\documentclass[a4paper,11pt]{article}
\usepackage{pos}

\newcommand{\dHybridR}{{\tt dHybridR}}

\newcommand{\rt}{R_{\rm tot}}
\newcommand{\trt}{\Tilde{R}_{\rm tot}}
\newcommand{\trs}{\Tilde{R}_{\rm sub}}
\newcommand{\rs}{R_{\rm sub}}

\newcommand{\w}[1]{v_{A,#1}}

\title{The Theory of Efficient Particle Acceleration at Shocks}
 \ShortTitle{Efficient DSA}

\author*[a,b]{Damiano Caprioli}
\author[c]{Colby Haggerty}
\author[d,e]{Pasquale Blasi}

\affiliation[a]{Department of Astronomy and Astrophysics, The University of Chicago, 5640 S Ellis Ave, Chicago, IL 60637, USA}
\affiliation[b]{Enrico Fermi Institute, The University of Chicago, 5640 S Ellis Ave, Chicago, IL 60637, USA}
\affiliation[c]{Institute for Astronomy, University of Hawaii, Honolulu, HI, 96822, USA}
\affiliation[d]{Gran Sasso Science Institute, Via F. Crispi 7, 67100 L' Aquila, Italy}
\affiliation[e]{INFN Laboratori Nazionali del Gran Sasso, Via G. Acitelli 22, Assergi, AQ, Italy}

\emailAdd{caprioli@uchicago.edu}

\abstract{The recent discoveries in the theory of diffusive shock acceleration (DSA) that stem from first-principle kinetic plasma simulations are discussed.
When ion acceleration is efficient, the back-reaction of non-thermal particles and self-generated magnetic fields becomes prominent and leads to both enhanced shock compression and particle spectra significantly softer than those predicted by the standard test-particle DSA theory. 
These results are discussed in the context of the non-thermal phenomenology of astrophysical shocks, with a special focus on the remnant of SN1006.
}

\FullConference{37$^{\rm{th}}$ International Cosmic Ray Conference (ICRC 2021)\\
		July 12th -- 23rd, 2021\\
		Online -- Berlin, Germany}

\begin{document}
\maketitle

%
%
%
\section{Introduction} \label{sec:intro}
Diffusive Shock Acceleration (DSA) is a prominent mechanism for producing relativistic particles  \cite{krymskii77,bell78a,blandford+78,axford+78}.
This special case of first-order Fermi acceleration, which applies to particles diffusing back and forth across the shock discontinuity, is particularly appealing because it naturally produces power-law spectra that depend only on the compression ratio, $r$,  i.e., the ratio of  downstream to upstream gas density. 
For a monoatomic gas with adiabatic index $\gamma=5/3$, the compression ratio and the spectral slope in momentum, $q$, read:
\begin{equation}\label{eq:qdsa}
    r=\frac{\gamma+1}{\gamma-1+2/M^2} \to 4; \qquad  q_{\rm DSA}= \frac{3r}{r-1}\to 4, 
\end{equation}
where we  took the limit of Mach number $M\gg1$ (strong shock).
Note that spectra are universal in momentum $p$ and scale as $f(p)\propto p^{-4}$, which means $f(E)\propto E^{-1.5}$ for non-relativistic particles and $f(E)\propto E^{-2}$ for relativistic cosmic rays (CRs).

DSA at supernova remnant (SNR) blast waves is widely considered the most promising way of producing the bulk of Galactic CRs, which requires the process to be inherently efficient: about $10-20\%$ of the shock kinetic energy must be converted in accelerated ions  \citep{baade+34,hillas05}.
Both observations of individual SNRs \citep{morlino+12, slane+14} and kinetic simulations \citep{caprioli+14a, caprioli+14b,park+15} support this scenario.

Nevertheless, when CRs carry a non-negligible fraction of the shock momentum/energy flux, they cannot be regarded as test-particles;
in  \emph{CR-modified shocks} both the shock dynamics and particle spectra must deviate from the standard predictions. 
The non-linear theory of DSA (e.g., \cite{jones+91,malkov+01} for reviews) suggests that the CR pressure should induce a shock precursor that leads to a total compression ratio $\rt\gg 4$, and thereby to CR spectra much flatter than the DSA prediction.
Accounting for the dynamical role of the  CR-driven magnetic turbulence limits such a compression to values $\lesssim 10$ \cite{caprioli+08, caprioli+09a}, but does not alter the theoretical expectation that strong shocks should be efficient accelerators and produce CR spectra with $q<4$.
However, such flat spectra are never observed in SNRs \citep{caprioli11,caprioli12}
or in radio SNe \cite{chevalier+06}, and would be inconsistent with the observed fluxes of Galactic CRs when propagation in the Milky Way is accounted for \cite{blasi+11a,evoli+19b}.
The reader is referred to \cite{caprioli15p,caprioli+19p,caprioli+20} and references therein for more details about the tension between DSA theory and observations, and for the possible solutions that have been suggested in the literature. 

In this work we present the results obtained with kinetic simulations of non-relativistic shocks \cite{haggerty+20,caprioli+20} that lay the groundwork for a theory of efficient DSA in which CR-modified shocks show both enhanced compression and steeper spectra with respect to the test-particle predictions.

\section{Hybrid Simulations of CR-modified shocks}
Hybrid simulations, with kinetic ions and neutralizing fluid electrons, are prominent tools for investigating CR acceleration and the production of magnetic turbulence. 
They have been used to perform a comprehensive analysis of ion acceleration at collisionless shocks as a function of the strength and topology of the pre-shock magnetic field, the nature of CR-driven instabilities, and the transport of energetic particles in the self-generated magnetic turbulence \cite{caprioli+14a,caprioli+14b,caprioli+14c}.  
Moreover, they have been used to unravel the processes that lead to the injection into DSA of protons \citep{caprioli+15}, ions with arbitrary mass/charge ratio \citep{caprioli+17}, and pre-existing CRs \citep{caprioli+18}. 

Here we use the code \dHybridR, a generalization of the classical hybrid approach that allows accelerated particles to become relativistic \cite{haggerty+19a}, to investigate the long-term evolution of non-relativistic shocks.
The reader can refer to \citep{haggerty+20,caprioli+20} for more information about the setup; 
one important ingredient is that we assumed the fluid electrons to be adiabatic, rather than prescribing an effective polytropic index $\gamma_e$ aimed to enforce electron/ion equipartition downstream (see appendix of \cite{caprioli+18}).
The latter choice, in fact, requires to fix the compression ratio a priori: if one guesses $r=4$, the electron equation of state becomes very stiff ($\gamma_e\sim 3$) and prevents any shock modification, enforcing $r\sim 4$ \cite{caprioli+14a}. 
Instead, using $\gamma_e\sim 5/3$, or iteratively setting $\gamma_e$ until equipartition is self-consistently achieved even for $r\gtrsim 4$, yields consistent results.

Our benchmark run considers a strong shock, with both sonic and Alfv\'enic Mach number $M=20$, propagating along a background magnetic field (parallel shock).
In this case, about 10\% of the shock kinetic energy is converted into accelerated particles \cite{caprioli+14a}.

\subsection{CR-induced Precursor and Postcursor}
\begin{figure}
\centering
\includegraphics[width=1\textwidth,clip=true,trim= 0 5 4 5]{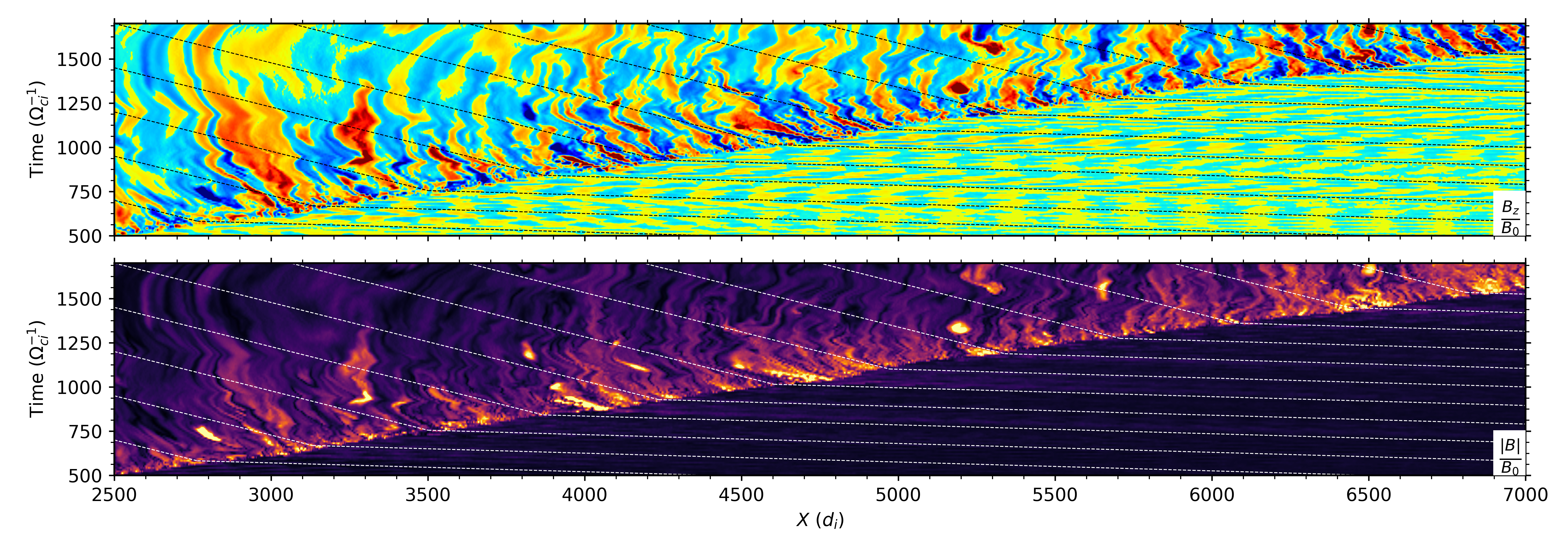}
\caption{Evolution of the magnetic field for our benchmark run \cite{haggerty+20}. 
At any given time, a 1D cut is plotted for $B_z$ and $|{\bf B}|$ (left and right).
The diagonal dashed lines correspond to average bulk flow plus the local Alfv\'en speed.
Immediately behind the shock, the magnetic structures align with such lines, which means that the phase motion of such structures points away from the shock.
}\label{fig:Bxt}
\end{figure}

Our benchmark run confirmed the prediction that, when DSA is efficient, the shock develops an upstream precursor, in which the incoming flow is slowed down and compressed under the effect of the CR pressure (see figure 2 of \cite{haggerty+20}).

What was unexpected is that the shock also develops a \emph{postcursor}, i.e., a region behind the shock where the dynamics is modified by the presence of CRs and self-generated magnetic perturbations. 
More precisely, we attest to the presence of an extended region in which magnetic structures drift at a finite speed \emph{towards downstream infinity} with respect to the thermal gas.
Both a visual inspection (see Fig.~\ref{fig:Bxt}) and a Fourier analysis (figure 7 of \cite{haggerty+20}) show that the phase speed of the magnetic fluctuations is comparable to the local Alfv\'en speed, both upstream and downstream;
as a result, CRs ---which tend to become isotropic in the wave frame---  also have a comparable net drift with respect to the background plasma (\cite{haggerty+20}, figure 5). 

The development of the postcursor implies that energy/pressure in CRs and magnetic fields are advected away from the shock at a faster rate that in a gaseous shock, which has has two crucial effects: 
1) it makes the shock behave as partially radiative, enhancing its compression;
2) it makes the CR spectrum steeper, enhancing the rate at which particles leave the acceleration region.

\subsection{Enhanced Shock Compression Ratio}
\begin{figure}[t]
\includegraphics[width=.45\textwidth,clip=true,trim= 1 0 0 0]{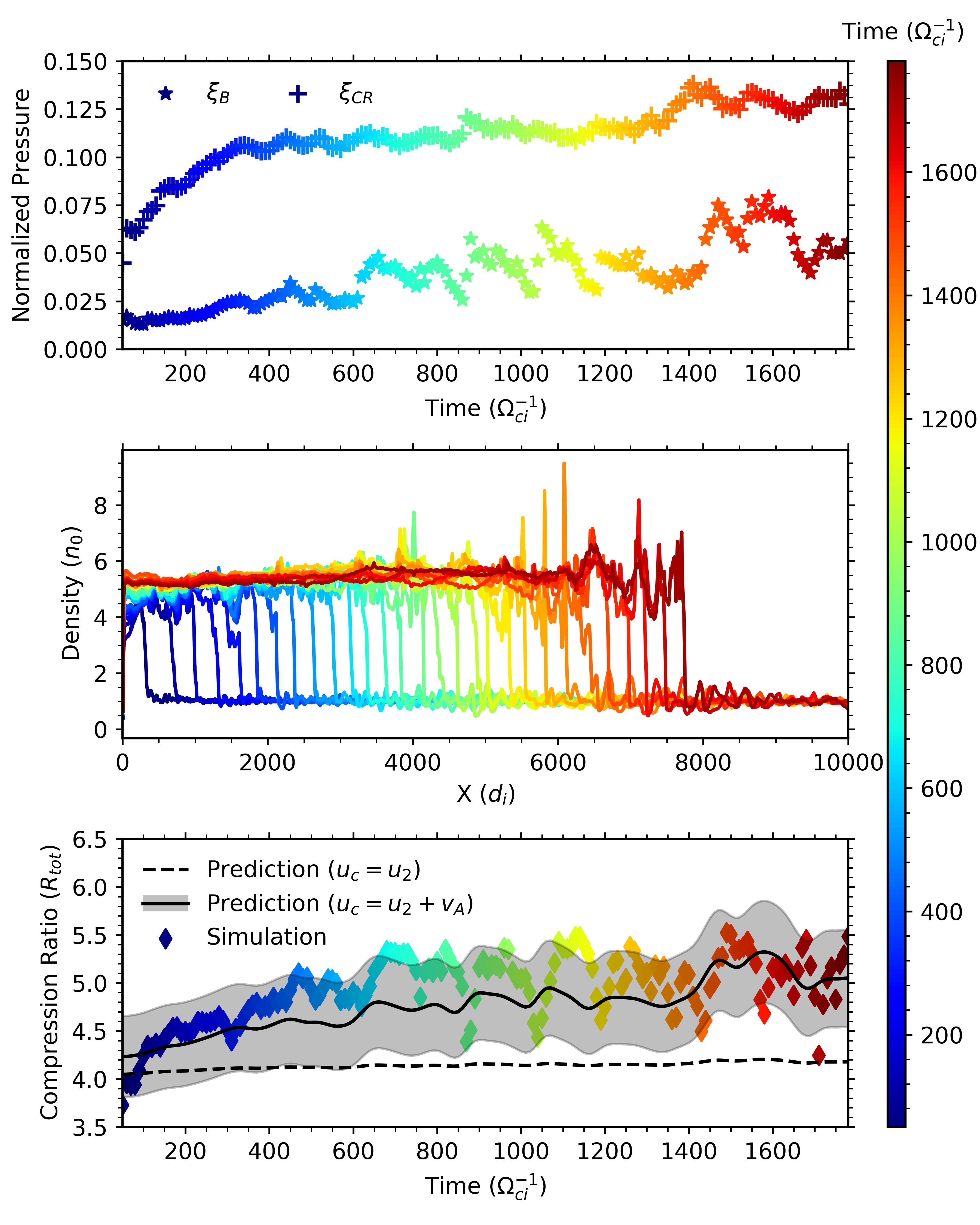}
\includegraphics[width=0.55\textwidth, clip=true,trim= 1 0 50 16]{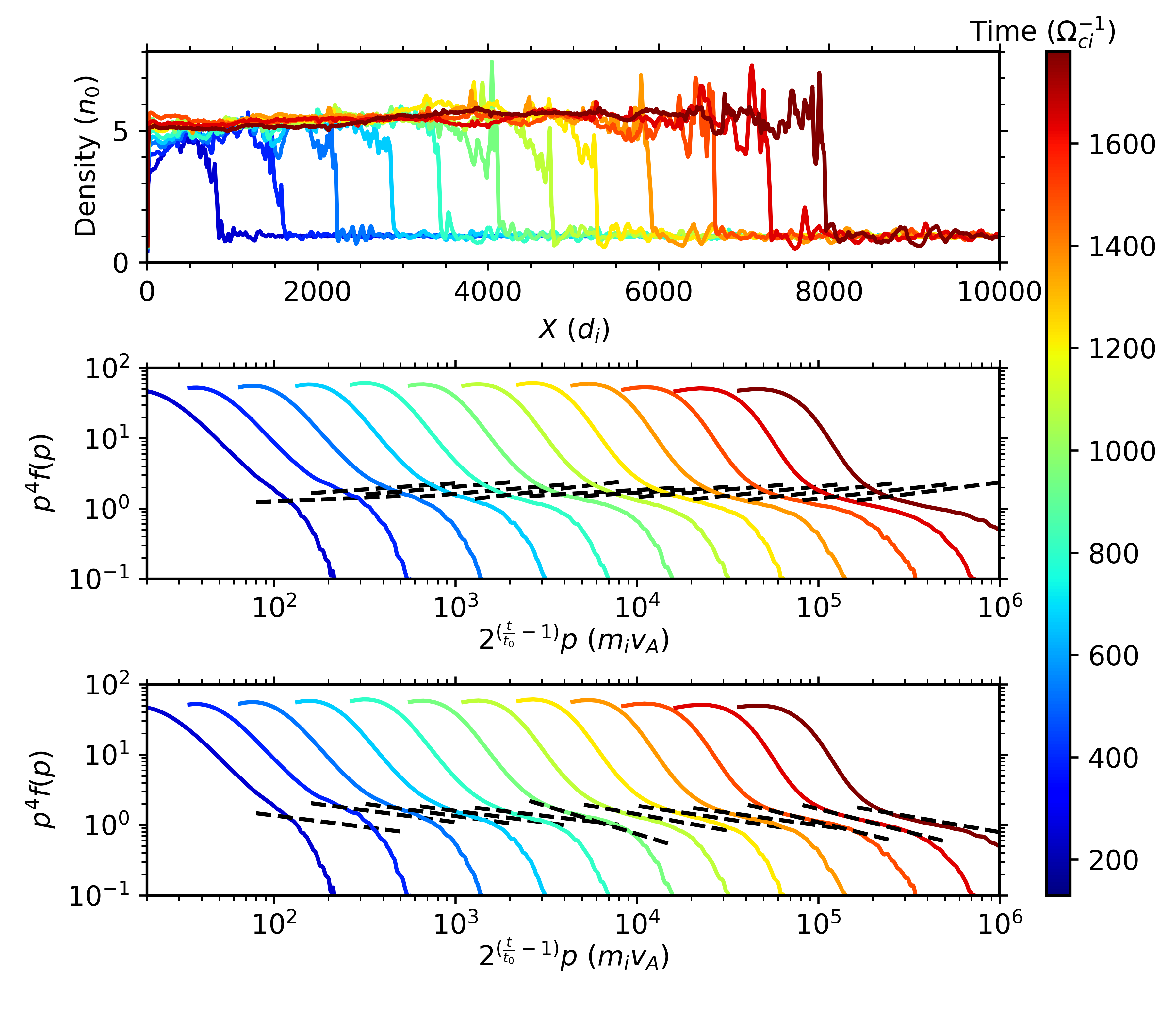}
\caption{
Time evolution (color coded) of physical quantities that show the CR-induced modification in our benchmark shock.
Left panels: normalized magnetic and CR pressures in the postcursor, $\xi_B$ and $\xi_c$  (top), density profile (middle), and total compression ratio, $\rt$ (bottom).
The CR pressure quickly converges to $\xi_c\approx 10\%$, while $\xi_B$ saturates around 6\%;
at the same time, the compression ratio departs from the test-particle value of $\sim 4$ and becomes $\rt\gtrsim 5.5$.
The prediction including the postcursor effect \cite{haggerty+20} is shown in gray.
Right panels: density profile (top) and post-shock particle spectra compared with standard (Equation \ref{eq:qdsa} with $r\to\rt$, middle) and revised prediction (Equation \ref{eq:qtilde}, bottom).
Despite the fluid shock compression ratio $\rt\to 6$, the CR non-thermal tail is significantly steeper than $p^{-4}$.}\label{fig:R}
\end{figure}

The left panels of Figure \ref{fig:R} illustrate the hydrodynamical modification induced by the postcursor. 
The normalized pressure in CRs and magnetic fields, $\xi_c$ and $\xi_B$, are plotted as a function of time in the first panel (crosses and stars, respectively);
the color code corresponds to the time in the simulation.
Together, the normalized CR and magnetic pressure encompass 15-20\% of the pressure budget in the postcursor:
$\xi_c$ increases quickly to a value $\gtrsim 0.1$ and remains nearly constant throughout the simulation, whereas the magnetic pressure rises more slowly up to $0.05-0.075$ towards the end of the simulation.
In appendix B of \cite{haggerty+20}, we solve the shock jump conditions between far upstream and the postcursor, including the contributions of CRs and Alfv\'en-like structures in the conservation of mass, momentum, and energy; 
such a solution accounts for the extra-compression observed in the simulation, where $\rt\to\sim 6$, as shown in the bottom left panel of Fig.~\ref{fig:R};
to stress the importance of the postcursor in the shock dynamics, such a panel also includes as a dashed line the prediction with no CR/magnetic drift. 
The CR pressure alone, without the magnetic/drift terms, is not sufficient to account for the strong shock modification that we observe, which demonstrates that the effect is inherently different from the enhanced compression expected in the classical theory of efficient DSA \cite{jones+91,malkov+01}.

\subsection{Steep Spectra}
Drastically different from the classical theory, an enhanced shock compression is not associated to flatter CR spectra, but rather to steeper ones, as shown in the right panels of Fig.~\ref{fig:R}. 
The standard prediction is that the CR momentum spectrum should flatten with time: such expected spectra would be described buy Eq.~\ref{eq:qdsa} with $r\to \rt$ and are shown with dashed lines in the middle panel.
It is straightforward to see that the measured postshock CR spectra (solid lines) are systematically steeper than such a prediction.
In the following, we label quantities at upstream infinity with $0$ and immediately upstream/downstream of the shock with 1,2.

Since CRs do not feel the change in speed of the thermal plasma, but rather that of their scattering centers, the effective compression ratio that they feel is:
\begin{equation}\label{eq:trt}
    \trt\simeq \frac{u_0}{u_2+\w2}\simeq \frac{\rt}{1 +\alpha}; \quad 
    \alpha\equiv \frac{\w2}{u_2},
\end{equation}
where in the numerator we set $\w0\approx 0$ because upstream infinity fluctuations should be small.
The $\alpha$ parameter quantifies the effect of the postcursor-induced spectral modification; 
since $\alpha>0$ (waves move towards downstream), the compression ratio felt by the CRs is always smaller than the fluid one.
At the same time, for low-energy CRs that probe the subshock only
\begin{equation}
    \trs\simeq \frac{u_1-v_{A,1}}{u_2+v_{A,2}}\simeq \rs\frac{1-\alpha_1}{1 +\alpha}; \quad 
    \alpha_1\equiv \frac{\w1}{u_1}.
\end{equation}
Note that, when the magnetic field is compressed at the shock and $B_2\approx\rs B_1$, we have $\alpha= \rs^{3/2}  \alpha_1\lesssim 8\alpha_1$ and the correction due to the postcursor dominates over the one in the precursor, which was the only effect accounted for in the previous literature \citep[e.g.,][]{zirakashvili+08b,caprioli12,kang+13}.

CR spectra turn out to be even steeper than $p^{-4}$ and match very well the slope calculated using $\trt$, namely
\begin{equation}\label{eq:qtilde}
    \Tilde{q}\equiv \frac{3\trt}{\trt-1}=
    \frac{3\rt}{\rt-1-\alpha}.
\end{equation}
This is plotted as dashed lines in the bottom right panel of Figure \ref{fig:R}.

In the elegant approach by Bell \citep[][]{bell78a}, power-laws arise as a balance between acceleration rate and escape;
following individual particles in the simulation, we show (figure 3 in \citep[][]{caprioli+20}) that the steepening is induced by an increase in the escape probability, rather than to a reduction of the acceleration rate \citep[or to effective losses due to the energy that CRs channel into waves, as suggested in ref.][]{bell+19}.
We have extensively checked this result against simulation parameters such as number of particles per cell, box transverse size, grid resolution, and time step choice; 
we rule out that the steep spectra are a numerical artifact to the best of our knowledge.

\section{A Revised Theory of Efficient DSA}
In summary, an interesting non-linear, physics-rich, picture arises. 
Quasi parallel shocks are inherently efficient in injecting thermal ions into the DSA mechanism \cite{caprioli+14a, caprioli+15};
DSA is self-sustaining and the streaming of energetic particles upstream of the shock triggers violent plasma instabilities (especially, the non-resonant one \citep{bell04}), which foster the rapid scattering and energization of CRs.
If no self-regulating effects kicked in, the DSA efficiency would grow uncontrolled, leading to the flatter and flatter spectra envisioned by the standard theory \cite{jones+91, malkov+01}.
Instead, when acceleration efficiency reaches $\sim 10-15\%$, the associated generation of magnetic field grows and back-reacts on both the shock modification and on the CR spectrum, as discussed above.
The net result of this non-linear chain is that both CR acceleration and $B$ amplification saturate, yielding a DSA efficiency of $\sim 10\%$ and CR spectra mildly steeper than $p^{-4}$.

One important implication is that the spectrum produced by efficient DSA is not universal, but rather depends on the strength of the self-generated  fields.
This adds a novel, crucial, physical constraint when modeling the non-thermal emission from CR sources, as outlined in \S6 of \cite{caprioli+20}.

This picture is indeed corroborated by multi-wavelength observations of given astrophysical objects (radio SNe and SNRs), as discussed in the contribution presented by R. Diesing at this conference \citep[also, see][]{diesing+21}.
In particular, SN1006 stands out as an ideal laboratory where to test these predictions because of its bilateral non-thermal emission, which is determined by the fact that its forward shock spans different shock inclinations \cite{reynoso+13}.
CR acceleration and shock modification are in fact predicted to occur only where the shock is quasi-parallel \citep{caprioli+14a}, which explains the strength and morphology of the observed X-ray synchrotron emission \cite{gamil+08} and the azimuthal modulation of the shock compression ratio [Giuffrida et al., submitted to Nature Communications].

\bibliography{Total_real}
\bibliographystyle{JHEP}

\end{document}